\documentclass[
preprint
]{ptephy_v1}
\usepackage{amsmath}
\usepackage{amssymb}
\usepackage{graphicx}
\usepackage{dcolumn}
\usepackage{bm}
\usepackage{hyperref}
\usepackage{microtype}

\preprintnumber{HUPD-2203}

\allowdisplaybreaks

\DeclareMathOperator{\diag}{diag}
\DeclareMathOperator{\rank}{Rank}
\DeclareMathOperator{\Tr}{Tr}

\begin{document}
\title{Renormalization group effects for a rank degenerate Yukawa matrix and the fate of the massless neutrino} 
\author{Nicholas J. Benoit}%
\affil{%
 Graduate School of Science, Hiroshima University,
Higashi-Hiroshima 739-8526, Japan
\email{d195016@hiroshima-u.ac.jp}}%
\author{Takuya Morozumi}%
\affil{%
Physics Program, Graduate School of Advanced Science and Engineering, \\
Hiroshima University, Higashi-Hiroshima 739-8526, Japan
}%
\affil{%
Core of Research for the Energetic Universe, Hiroshima University, \\
Higashi-Hiroshima 739-8526, Japan
\email{morozumi@hiroshima-u.ac.jp}
\email{yu-shimizu@hiroshima-u.ac.jp}}%
\author[2]{Yusuke~Shimizu}%
\author{Kenta Takagi}%
\affil{%
Higashi-Hiroshima, Hiroshima Prefecture, Japan
}%
\author{Akihiro Yuu}%
\affil{%
Yokkaichi, Mie Prefecture, Japan
}%

\subjectindex{xxxx, xxx}

\begin{abstract}
The Type-I seesaw model is a common extension to the Standard Model that describes neutrino masses.
The Type-I seesaw introduces heavy right-handed neutrinos with Majorana mass that transform as Standard Model electroweak gauge singlets.
We initially study a case with two right-handed neutrinos called the 3-2 model.
At an energy scale above the right-handed neutrinos, the effective neutrino mass matrix is rank degenerate implying the lightest neutrino is massless.
After considering renormalization effects below the two right-handed neutrinos, the effective neutrino mass matrix remains rank degenerate.
Next, we study a model with three right-handed neutrinos called the 3-3 model.
Above the energy scale of the three right-handed neutrinos, we construct the effective neutrino mass matrix to be rank degenerate.
After solving for the renormalization effects to energies below the three right-handed neutrinos, we find the rank of the effective neutrino mass matrix depends on the kernel solutions of the renormalization group equations.
We prove for the simplest kernel solutions the effective neutrino mass matrix remains rank degenerate.
\end{abstract}

\maketitle
\section{Introduction}
The knowledge of the absolute masses of neutrinos is still lacking despite
the precise determination of mass squared differences from neutrino oscillation experiments\cite{Esteban:2020cvm,NuFit:2021,Aker:2021gma,Planck:2018vyg}.
If the neutrino mass type is Majorana, the lightest absolute mass is a source of the uncertainty in the determination of the rate of neutrinoless double beta decay\cite{Furry:1939qr,Schechter:1981bd,Nieves:1984sn,Takasugi:1984xr,Dolinski:2019nrj}.
Actuality, there are models that allow for the lightest neutrino to be massless\cite{Ma:1998zg}.
For those models, the neutrinoless double beta decay rate becomes more predictive since the non-zero masses are directly obtained from the mass squared differences and only a single Majorana phase is allowed.
In cosmology, the absolute neutrino masses are important to early universe phenomena such as big-bang nucleosynthesis, and large scale structure formation\cite{Dolgov:2002wy}.
And we have previously discussed the role of the absolute masses for neutrinos in the time evolution of lepton number\cite{Adam:2021qiq,Adam:2021yvr}.
Thus, understanding the absolute masses of neutrinos is important for neutrinoless double beta decay, cosmology, and the time evolution of lepton number.
The main goal of this paper is to study a massive neutrino models' stability, with the lightest neutrino being massless, including 1-loop quantum corrections.

We consider the massive neutrino model called the Type-I seesaw \cite{Mohapatra:1979ia,Minkowski:1977sc,Gell-Mann:1979vob,Glashow:1979nm,Yanagida:1979as} first with two right-handed neutrinos, the 3-2 seesaw model; then with three right-handed neutrinos, the 3-3 seesaw model\footnote{A model with only one right-handed fermion is excluded by the measured mass squared differences of oscillation experiments.}.
For the 3-2 seesaw model the neutrino Dirac mass matrix is a 3 by 2 matrix and its rank is two.
The 3-2 seesaw model has been previously studied sometimes under the name the minimal Type-I seesaw model \cite{Ma:1998zg,King:1999mb,King:2002nf,Frampton:2002qc,
Endoh:2002wm,Shimizu:2017fgu,Xing:2020ald,Zhou:2021bqs}.
For the model with three right-handed neutrinos, we assume a texture of the neutrino Dirac mass terms to reduce the rank of the matrix to two.
Both of those models lead to a vanishing mass for the lightest neutrino, i.e., $m_1=0$ for normal hierarchy.
In the usual Type-I seesaw the right-handed neutrinos have a significantly heavier mass than the Standard Model particle content to reproduce small neutrino masses.
This means the heavy right-handed neutrinos are integrated out at low-energy scales\footnote{Different nomenclature is used in literature to name the heavy right-handed neutrinos, for example right-handed neutrinos or heavy neutrino singlets or simply heavy neutrinos}.

In this paper, we integrate out the right-handed neutrinos within the tree level matching approximation and take into account the one loop renormalization group (RG) effects\cite{Antusch:2002rr,Antusch:2005gp}.
Then we determine the matrix rank of the effective neutrino mass matrix.
The renormalization group running of neutrino masses has been studied previously in literature with the rank of the 3-2 seesaw model being considered for the one loop matching approximation \cite{Zhou:2021bqs} and two loop renormalization group effects\cite{Davidson:2006tg}.
However, we believe the stability of the 3-3 seesaw model with a matrix of rank two has not been examined against renormalization group effects.

We organize this paper as follows.
In section \ref{sec:TypeISee}, we introduce the seesaw model as a review.
In section \ref{sec:MatrixRank}, we demonstrate the method we use to determine the rank of a matrix.
Then in section \ref{sec:32Seesaw}, we study the renormalization group effects in the 3-2 seesaw model and calculate the rank of the effective Majorana mass matrix.
Next in section \ref{sec:33Seesaw}, we apply the techniques of section \ref{sec:32Seesaw} for the 3-3 seesaw model with generic solutions to the renormalization group equations.
Then using specific renormalization group solutions, we consider the rank of the effective mass matrix in section \ref{sec:33SeesawRank}.
Lastly, section \ref{sec:Conclusion} is devoted to summary and discussion.
Appendix \ref{sec:RGE} holds the details of the renormalization group equations and solutions used in the main text.

\section{Type-I seesaw model}\label{sec:TypeISee}
\begin{figure}[htb]
\centering
\includegraphics[width=400pt]{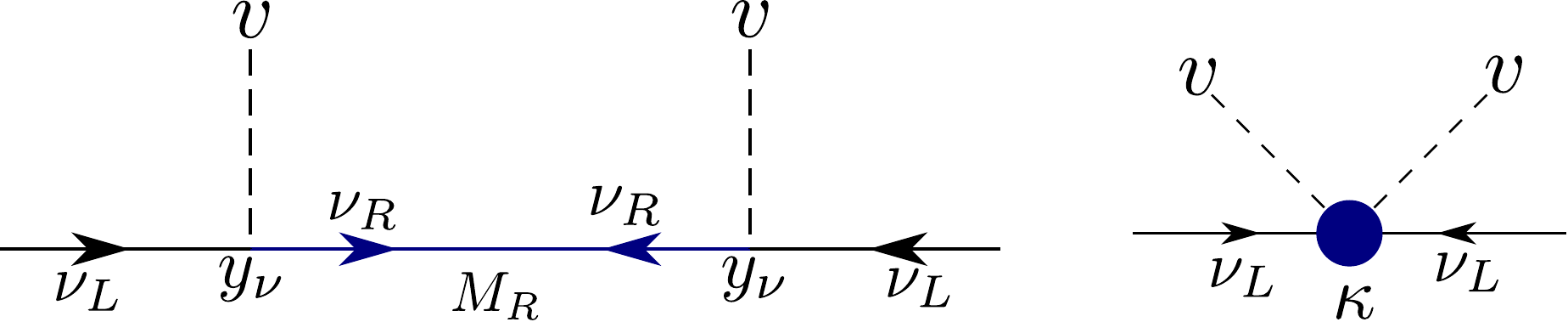}
\caption{\label{Fig:SeesawDiagrams}Diagrams for the Type-I seesaw model at different energy scales.
The left diagram is for the full theory and the right diagram is for the effective theory. By integrating out the right-handed neutrinos shown in the left diagram, the coefficient of the Weinberg operator in the right diagram is generated.}
\end{figure}

We consider a general Type-I seesaw model with right-handed neutrinos $N_n$, the standard model lepton doublets $l_L$, and the standard model Higgs doublet $\widetilde{\phi}=i\tau_2\phi^\ast$ as an extension to the Standard Model \cite{Mohapatra:1979ia,Minkowski:1977sc,Gell-Mann:1979vob,Glashow:1979nm,Yanagida:1979as}.
The Lagrangian of the extended section reads,
\begin{equation}
 \mathcal{L}_N=\frac{1}{2}\overline{N_n} i\slash\mspace{-9.0mu}\partial N_n -\frac{1}{2}\overline{N_n}M_{R\,n}N_n- (y_{\nu j n}\overline{l_{j L}}\widetilde{\phi}N_n + \text{h.c.}),
 \label{Eq:Lagrangian}
\end{equation}
where $N_n^C=N_n$  and  $C$ denotes charge conjugation, and $n$ are integer values that correspond to the number of the heavy right-handed neutrinos.
Under the Lagrangian of Eq.(\ref{Eq:Lagrangian}) the masses of the active neutrinos are found by diagonalizing the effective Majorana mass matrix,
\begin{equation}
 m_{eff}=-\frac{v^2}{2}y_{\nu}\frac{1}{M_{R}}y^T_{\nu},
 \label{eq:effectivemassmatrix}
\end{equation}
where $v=246$ GeV is the Higgs vacuum expectation value.
If we neglect quantum effects captured by running the renormalization group equations, $m_{eff}$ does not change with the energy scale.
However, if we want to reproduce the masses of the active neutrinos to be in the eV region, as constrained by experiments, the masses of $M_R$ are required to be near the GUT scale.
This means for practical experiments the right-handed neutrinos are integrated out, down to the energy scale being considered \cite{Antusch:2002rr,King:2000hk}.
After integration, the Wilson coefficient of the so-called Weinberg operator, $\kappa(\mu)$, is generated \cite{Weinberg:1979sa}.
\begin{equation}
  \mathcal{O}_5 = \frac{1}{4} \kappa_{gh}^{\ast}(\mu)\left(\overline{l^C_L}^{\,g}\cdot\phi\right) \left(l^h_L\cdot\phi\right)+\text{h.c.}
\end{equation}
As the operator is dimension five, it is suppressed by the masses of the right-handed neutrinos $M_R$.
We display the diagrams representing how the Wilson coefficient of the Weinberg operator appears after integration in figure \ref{Fig:SeesawDiagrams}.

The quantum effects captured by the renormalization group equations mean the effective mass matrix of Eq.(\ref{eq:effectivemassmatrix}) becomes energy scale dependent $m_{eff}(\mu)$.
In the Type-I seesaw model, the energy scale dependence is realized with the equation,
\begin{equation}
  \overset{(a)}{m}_{eff}(\mu)=-\frac{v^2}{4}\overset{(a)}{\kappa}(\mu)-\frac{v^2}{2}\overset{(a)}{y_{\nu}}(\mu)\frac{1}{\overset{(a)}{M_{R}}(\mu)}\overset{(a)}{y_{\nu}}(\mu)^T
 \label{Eq:EffectiveMassmu}
\end{equation}
where $\kappa(\mu)$, $y_\nu(\mu)$, and $M_R(\mu)$ depend on the energy scale $\mu$ being calculated up to a cutoff scale $\Lambda$\footnote{We do not consider what occurs for scales above $\Lambda$ to generate mass for the right-handed neutrinos.}.
We have adopted the superscript notation to denote the different renormalization energy regions from the work of Antusch \cite{Antusch:2005gp}, for which the lowest region is denoted as $a=1$.

\section{Determination of the rank of the mass matrix}\label{sec:MatrixRank} 
In this section, we present the method we use to determine the rank of the effective Majorana mass matrix of Eq.(\ref{Eq:EffectiveMassmu}).
Precisely speaking, we are interested in the rank of the Hermite matrix that determines the mass squared eigenvalues for light neutrinos,
\begin{equation}
  \overset{(1)}{\mathcal{M}} = \overset{(1)}{m}_{eff}(\mu_1)[\overset{(1)}{m}_{eff}(\mu_1)]^\dagger,
  \label{eq:hermitemeff}
\end{equation}
where $\overset{(1)}{m}_{eff}$ denotes the effective Majorana mass matrix of Eq.(\ref{Eq:EffectiveMassmu}) for the light neutrinos obtained after integrating out all the right-handed neutrinos.
The $\mu_1$ is the mass scale of the lightest right-handed Majorana neutrino.
We can solve for the rank of the Hermite matrix using the characteristic polynomial,
\begin{equation}
  \det{(\overset{(1)}{\mathcal{M}}-x)} = 0.
\end{equation}
The eigenvalue $x$ satisfies the following equation,
\begin{equation}
  -x^3+x^2 I_2-x I_1 +I_0=0,
  \label{eq:characteristicpolynomial}
\end{equation}
where $I_0,I_1,\text{ and }I_2$ are calculated using the Faddeev-LeVerrier algorithm,
\begin{align}
  I_2 & = \text{Tr}[\overset{(1)}{\mathcal{M}}], \label{eq:invariant2}\\
  I_1 & = \frac{1}{2} \left(
    (\text{Tr}[\overset{(1)}{\mathcal{M}}])^2
    -\text{Tr}[(\overset{(1)}{\mathcal{M}})^2] \right),\label{eq:invariant1} \\
  \begin{split}
  I_0 & = \frac{1}{6} \left(
    (\Tr[\overset{(1)}{\mathcal{M}}])^3
    -3\Tr[\overset{(1)}{\mathcal{M}}]
    \Tr[(\overset{(1)}{\mathcal{M}})^2]+2\Tr[(\overset{(1)}{\mathcal{M}})^3]\right)=\det \overset{(1)}{\mathcal{M}}.
  \end{split}
  \label{eq:invariant0}
\end{align}
Since the characteristic polynomial is invariant under the unitary transformation $\mathcal{M}^\prime=U^\dagger \mathcal{M} U$, the coefficients; $I_0$, $I_1$, and $I_2$, are also invariants.

In order to prove the rank explicitly, we calculate the matrix invariants $I_0$, $I_1$, and $I_2$ of the Hermite matrix in Eq.(\ref{eq:hermitemeff}) using Eqs.(\ref{eq:invariant2}-\ref{eq:invariant0}).
Then rank of Eq.(\ref{eq:hermitemeff}) is determined by subtracting the degree of the polynomial with the superscript of the 0 root in Eq.(\ref{eq:characteristicpolynomial})\footnote{We are assuming the Hermite matrix is diagonalizable.}.
If $I_0=0$ and the other invariants are non-zero the rank of Eq.(\ref{eq:hermitemeff}) is $3-1=2$, and one of the masses squared is zero with the others being massive.
This is clear when the invariants are written in terms of the eigenvalues of the Hermite matrix
\begin{align}
  \label{eq:invariant2mass}
  I_2&=m_1^2+m_2^2+m_3^2,
  \\
  I_1&=m_1^2m_2^2+m_2^2m_3^2+m_1^2m_3^2,
  \\
  I_0&=(m_1m_2m_3)^2.
  \label{eq:invariant0mass}
\end{align}
Then, for example, to have $I_0=0$ any of the eigenvalues can equal zero, but the other two eigenvalues must remain non-zero to have $I_1$ and $I_2$ as non-zero.

\section{3-2 Type-I seesaw model renormalization group running}\label{sec:32Seesaw}
\begin{figure}[htb]
 \centering
 \includegraphics[width=125mm]{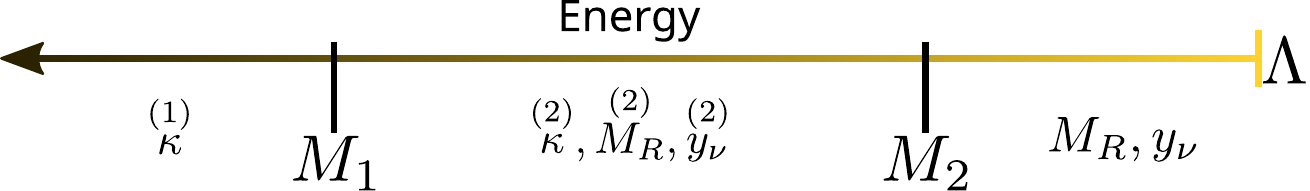}
  \caption{\label{Fig:2EnergyScales}
 Illustration of the different energy regions being considered in the 3-2 Type-I seesaw model.
 The highest energy region occurs above $M_2$ where no $\kappa$ term exists.
 The renormalization energy scale decreases from right to the left.
 The matrices $\kappa(\mu)$, $y_\nu(\mu)$, and $M_R(\mu)$ are scale dependent based on the $\beta$-functions $\beta_\kappa$, $\beta_{y_\nu}$, and $\beta_{M_R}$ defined by the corresponding renormalization group equations.
 To denote the different renormalization energy regions we adopt the superscript notation from \cite{Antusch:2005gp}.}
\end{figure}

In this section, we solve the renormalization group equations for two different energy regions of the 3-2 Type-I seesaw model.
The 3-2 Type-I seesaw model is the minimalist version of the Type-I seesaw that is not directly excluded by neutrino experiments and is sometimes named the minimal Type-I seesaw model (MSM)\cite{Ma:1998zg,King:1999mb,King:2002nf,Frampton:2002qc,Endoh:2002wm,Shimizu:2017fgu,Xing:2020ald,Zhou:2021bqs}.
This version introduces two new right-handed neutrinos to the Standard Model that we label as $N_1 $ and $N_2$ following Eq.(\ref{Eq:Lagrangian}).
We assume the masses of the right-handed neutrinos are hierarchical, satisfying $M_1 < M_2 < \Lambda$ according to figure \ref{Fig:2EnergyScales}.
Consequently, the renormalization scales are defined as;
\begin{equation}
  t_1 \equiv \log \frac{M_2}{M_1}, \quad t \equiv \log \frac{M_2}{\mu}.
\end{equation}
At the initial renormalization scale of $t=0\;(\mu=M_2)$, we go to the diagonal basis for the right-handed mass matrix,  
\begin{equation}
  M_R(0)=\diag(M_1(0),\,M_2(0)).
\end{equation}
In this basis, the neutrino Yukawa matrix has a $3 \times 2$ structure,
\begin{equation}
  y_{\nu}(0) = \begin{pmatrix}
    y_{\nu 1}(0) & y_{\nu 2}(0)
  \end{pmatrix}.
  \label{Eq:HighEnergyYukawa} 
\end{equation}
An important feature of the 3-2 Type-I seesaw occurs at or below the initial renormalization scale of $t=0$.
There only $M_R(0)$ and $y_{\nu}(0)$ contribute to the effective neutrino mass calculation using Eq.(\ref{Eq:EffectiveMassmu}).
The resulting $3\times3$ effective mass matrix $m_{eff}(0)$ is rank degenerate, i.e. $\rank{m_{eff}(0)}=2$; which, based on the discussions of Eqs.(\ref{eq:invariant2mass}-\ref{eq:invariant0mass}), means the lightest neutrino mass eigenvalue is zero.

Slightly above $t=0$, the renormalization energies are smaller than the mass of the right-handed neutrino $M_2$.
Thus, we must integrate out the right-handed neutrino $N_2$, which generates the following Wilson coefficient,
\begin{equation}
 \overset{(2)}{\kappa}(0) = 2y_{\nu 2}(0)\frac{1}{M_2(0)}y_{\nu 2}^T(0)
 \label{Eq:Weinberg32-1}
\end{equation}
We use the usual method of tree-level matching to demand the effective masses of the neutrinos $m_{eff}(\mu)$ are continuous over the energy regions.
Specifically, we demand that the $m_{eff}(\mu)$'s are equal before and after $N_2$ is integrated out;
\begin{equation}
 \overset{(2)}{m_{eff}}(0) =m_{eff}(0), 
  \label{Eq:Treelevelmatch}
\end{equation}
where from Eq.(\ref{Eq:EffectiveMassmu}) and Eq.(\ref{Eq:Weinberg32-1}) we have,
\begin{align}
   m_{eff}(0) =& -\frac{v^2}{2}y_{\nu}(0)\frac{1}{M_{R}(0)}y^T_{\nu}(0), \\
   \overset{(2)}{m_{eff}}(0) =& -\frac{v^2}{4}\overset{(2)}{\kappa}(0)
    -\frac{v^2}{2}\overset{(2)}{y_{\nu}}(0)\left(\overset{(2)}{M_R}(0)\right)^{-1}\left(\overset{(2)}{y_{\nu}}(0)\right)^T, 
\end{align}
where $\overset{(2)}{M_R}(0)={M_1}(0)$.
The Wilson coefficient $\overset{(2)}{\kappa}(t)$, the neutrino Yukawa matrix $\overset{(2)}{y_{\nu}}(t)$, and the remaining right-handed mass $\overset{(2)}{M_R}(t)$ are energy scale dependent and are only valid for the scales $ 0 < t < t_1 $, see figure \ref{Fig:2EnergyScales}.
We calculate the corrections from the energy scale dependence with the one-loop renormalization group equations of Appendix \ref{sec:RGE}.

As an example, we write the one-loop renormalization group equation for $\overset{(2)}{\kappa}(t)$;
\begin{equation}
  -16 \pi^2 \frac{d\overset{(2)}{\kappa}(t)}{dt} =
    \left(\overset{(2)}{y_{\nu}}(t)\overset{(2)}{y_{\nu}}(t)^{\dagger}\right) \overset{(2)}{\kappa}(t) +
    \overset{(2)}{\kappa}(t) \left(\overset{(2)}{y_{\nu}}(t)\overset{(2)}{y_{\nu}}(t)^{\dagger}\right)^T +
    \overset{(2)}{\alpha_\kappa}(t) \overset{(2)}{\kappa}(t),
\end{equation}
which has a solution at boundary of the lower energy scale $t_1$ given by,
\begin{equation}
  \overset{(2)}{\kappa}(t_1) = e^{-\frac{1}{16\pi^2} \int_0^{t_1} ds\, \overset{(2)}{\alpha_\kappa}(s)}
  \overset{(2)}{W}(t_1) \overset{(2)}{\kappa}(0) \overset{(2)}{W}(t_1)^T,
  \label{eq:KappaRGESolution32}
\end{equation}
where we define the solution kernel $\overset{(2)}{W}(t_1)$ following Eq.(\ref{eq:Kernelkappa}),
\begin{equation}
  \overset{(2)}{W}(t_1) \equiv T \exp\left[-\frac{1}{16\pi^2}\int_0^{t_1}(\overset{(2)}{y_{\nu}}(s)\overset{(2)}{y_{\nu}}(s)^{\dagger}) ds \right].
  \label{eq:KappaKernel}
\end{equation}
The $T$ denotes the exponential is ordered by decreasing energy scale.
The energy dependence for the terms $\overset{(2)}{y_{\nu}}(t)$ and $\overset{(2)}{M_R}(t)$ are solved similarly using Eqs.(\ref{eq:YukawaRGESolution}-\ref{eq:RightMassRGESolution}).
Now we have solutions for $\overset{(2)}{\kappa}(t_1)$, $\overset{(2)}{y_{\nu}}(t_1)$, and $\overset{(2)}{M_R}(t_1)$ down to the energy scale $t=t_1$, which is equivalent to saying $\mu=M_1$.
At this scale we must integrate out the remaining right-handed neutrino $N_1$ and take a tree-level matching,
\begin{align}
  \overset{(1)}{m_{eff}}(t_1)&=\overset{(2)}{m_{eff}}(t_1)
\\
  \overset{(1)}{m_{eff}}(t_1)&=-\frac{v^2}{4}\overset{(2)}{\kappa}(t_1)
  -\frac{v^2}{2}\overset{(2)}{y_{\nu}}(t_1)\left(\overset{(2)}{M_R}(t_1)\right)^{-1}\left(\overset{(2)}{y_{\nu}}(t_1)\right)^T,
  \label{eq:32treematch1}
\end{align}
that is similar to Eq.(\ref{Eq:Treelevelmatch}).

In contrast to the scale of $t=0$, for the scales $t>t_1$ we have integrated out all right-handed neutrinos.
Thus, the neutrino effective mass matrix only depends on the Wilson coefficient,
\begin{equation}
 \overset{(1)}{m_{eff}}(t_1) = -\frac{v^2}{4} \overset{(1)}{\kappa}(t_1).
\end{equation}
We could solve the renormalization group equation for $\overset{(1)}{\kappa}(t_1)$ down to energy scales of practical experiments and apply the solution to different observables.
However, above $t=t_1$, or below $M_1$, the one-loop renormalization group equation of $\overset{(1)}{\kappa}(t)$ from Eq.(\ref{eq:KappaRGE}) can not change its rank\footnote{See section 3.3 of \cite{Antusch:2005gp} for a more detailed analysis on why this is the case}.
Therefore, we focus on the rank of $\overset{(1)}{\kappa}(t_1)$ to study if the matrix rank of $\overset{(1)}{m_{eff}}(t_1)$ is equal to the matrix rank of the high energy $m_{eff}(0)$.

To determine the rank of the low energy mass matrix $\overset{(1)}{m_{eff}}(t_1)$ we investigate the parts of the Wilson coefficient $\overset{(1)}{\kappa}(t_1)$.
The parts of $\overset{(1)}{\kappa}(t_1)$ are determined from the tree-level matching of Eq.(\ref{eq:32treematch1}) resulting in,
\begin{equation}
 \overset{(1)}{\kappa}(t_1) =2  \overset{(2)}{y_{\nu}}(t_1)\left(\overset{(2)}{M_R}(t_1)\right)^{-1}\overset{(2)}{y_{\nu}}(t_1)^{T}+\overset{(2)}{\kappa}(t_1).
\end{equation}
Then substituting the solutions to the renormalization group equation of Eq.(\ref{eq:KappaRGESolution32}) for $\overset{(2)}{\kappa}(t_1)$;
\begin{equation}
 \overset{(1)}{\kappa}(t_1) =2  \overset{(2)}{y_{\nu}}(t_1)\left(\overset{(2)}{M_R}(t_1)\right)^{-1}\overset{(2)}{y_{\nu}}(t_1)^{T}
 +e^{-\frac{1}{16\pi^2} \int_0^{t_1} \overset{(2)}{\alpha_\kappa}(s) ds}
  \overset{(2)}{W}(t_1) \overset{(2)}{\kappa}(0) \overset{(2)}{W}(t_1)^T.
\end{equation}
At this point we note two important considerations.
First, the contribution from the running of the right-handed mass $\overset{(2)}{M_R}(t_1)$ is a scaling factor that will not affect the rank of the low energy mass matrix.
Second, the running of the Yukawa couplings are what determines the rank of $\overset{(1)}{m_{eff}}(t)$ for reasons to be explained.
We explicitly write the Yukawa couplings in terms of the solutions to the renormalization group equations up to the energy scale $\mu=M_2 $ or $t=0$.
\begin{equation}
   \overset{(2)}{y_{\nu}}(t_1) =
     e^{-\frac{1}{16\pi^2}\int^{t_1}_0 \overset{(2)}{\alpha_{y_\nu}}(s)ds}
   \overset{(2)}{U}(t_1) \overset{(2)}{y_{\nu}}(0)
   \label{eq:32YukawaSolution}
\end{equation}
where the kernel function $\overset{(2)}{U}(t_1)$ is the solution to the Yukawa renormalization group equation of Eq.(\ref{eq:YukawaRGE}).
Note, a subtle difference between the kernel of Eq.(\ref{eq:Kernelkappa}) for $\overset{(2)}{\kappa}(t_1)$ and the kernel of Eq.(\ref{eq:Kernelyukawa}) for $\overset{(2)}{y_{\nu}}(t_1)$ is an additional $\frac{3}{2}$ factor in the exponential.
Then, we write the matrix $\overset{(1)}{m_{eff}}(t_1)=\overset{(1)}{\kappa}(t_1)$ with the two kernel functions from Eq.(\ref{eq:32YukawaSolution}) and Eq.(\ref{Eq:Weinberg32-1}),
\begin{equation}
 \begin{split}
    \overset{(1)}{\kappa}(t_1) &=2 e^{-\frac{1}{8\pi^2}\int^{t_1}_0 \overset{(2)}
      {\alpha_{y_\nu}}(s) ds}
      \left(\overset{(2)}{M_R}(t_1)\right)^{-1}
      \overset{(2)}{U}(t_1) y_{\nu 1}(0) y_{\nu 1}^T(0) \overset{(2)}{U}(t_1)^T\\
    &\quad +2e^{-\frac{1}{16\pi^2} \int_0^{t_1} \overset{(2)}{\alpha_\kappa}(s) ds}
      \frac{1}{M_2(0)}
      \overset{(2)}{W}(t_1) y_{\nu 2}(0)y_{\nu 2}^T(0) \overset{(2)}{W}(t_1)^T,
 \end{split} \label{eq:32effectiveRank}
\end{equation}
where we have also used Eq.(\ref{Eq:HighEnergyYukawa}) to write the Yukawa vectors explicitly.

Next, we focus on the relationship between the kernel functions and the Yukawa vectors.
In Eq.(\ref{eq:32simpleeffectiveRank}) the two kernel functions appear as $\overset{(2)}{W}(t_1)$ and $\overset{(2)}{U}(t_1)$, and are square matrices shaped by the number of rows in the Yukawa vectors $y_{\nu 1}(0)$ and $y_{\nu 2}(0)$.
Upon multiplication, the kernel functions act, separately, as linear transformation matrices on the Yukawa vectors.
To better illustrate the rotation we rewrite the effective mass matrix of Eq.(\ref{eq:32effectiveRank}) to be,
\begin{equation}
 \overset{(1)}{m_{eff}}(t_1)
 =  X \,\overset{(2)}{U}(t_1) y_{\nu 1}(0) y_{\nu 1}^T(0) \overset{(2)}{U}(t_1)^T
 + Z \, \overset{(2)}{W}(t_1) y_{\nu 2}(0)y_{\nu 2}^T(0) \overset{(2)}{W}(t_1)^T,
 \label{eq:32simpleeffectiveRank}
\end{equation}
where $X$ and $Z$ are scaling functions that do not contribute to the matrix rank.
They are given as,
\begin{gather}
    X \equiv -\frac{v^2}{2} e^{-\frac{1}{8\pi^2}\int^{t_1}_0 \overset{(2)}{\alpha_{y_\nu}}(s) ds}\left(\overset{(2)}{M_R}(t_1)\right)^{-1} \\
    Z \equiv -\frac{v^2}{2} e^{-\frac{1}{16\pi^2} \int_0^{t_1} \overset{(2)}{\alpha_\kappa}(s) ds}\frac{1}{M_2(0)}.
\end{gather}
Thus, we rewrite Eq.(\ref{eq:32simpleeffectiveRank}) after the linear transformations have occurred,
\begin{equation}
 \overset{(1)}{m_{eff}}(t_1)
   =X \,y^U_{\nu 1}(t_1) y^U_{\nu 1}(t_1)^T+Z \, y^W_{\nu 2}(t_1)y^W_{\nu 2}(t_1)^T,
 \label{eq:TransformedMass32}
\end{equation}
where the transformed vectors of,
\begin{gather}
    y^U_{\nu 1}(t_1)=\overset{(2)}{U}(t_1) y_{\nu 1}(0) \label{eq:32transformedYukawa1}\\
    y^W_{\nu 2}(t_1)=\overset{(2)}{W}(t_1) y_{\nu 2}(0) \label{eq:32transformedYukawa2}
\end{gather}
are non-zero and assumed to be linearly independent of each other. 
With the effective mass matrix given in Eq.(\ref{eq:TransformedMass32}), the mass squared eigenvalues of the Hermite matrix $\overset{(1)}{\mathcal{M}}=\overset{(1)}{m_{eff}}(t_1) [\overset{(1)}{m_{eff}}(t_1)]^\dagger$ has a zero eigenvalue.  
To prove this, we evaluate three invariants in Eqs.(\ref{eq:invariant2}-\ref{eq:invariant0}) with the effective Majorana mass matrix for 3-2 model in Eq.(\ref{eq:TransformedMass32}).
The first two invariants $I_1\neq0$ and $I_2\neq0$ hold the expressions below,
\begin{align}
  \begin{split}\label{eq:32I1result}
    I_1 = & |X|^2 |Z|^2 \left\{ y^{W\ast}_{12} \left[(y^W_{12} y^U_{21}-y^U_{11}y^W_{22}) 
      y^{U\ast}_{21}+(y^W_{12} y^U_{31}-y^U_{11}y^W_{32})y^{U\ast}_{31}\right]\right. \\
    & \left.+y^{U\ast}_{11}\left[(y^U_{11}y^W_{22}-y^W_{12}y^U_{21})y^{W\ast}_{22}+(y^U_{11} 
      y^W_{32}-y^W_{12} y^U_{31})y^{W\ast}_{32}\right]+|y^W_{22} y^U_{31}-y^U_{21}y^W_{32}|^2 \right\}^2,
  \end{split}
\\
  \begin{split}\label{eq:32I2result}
    I_2 = & |X (y^U_{11})^2+Z (y^W_{12})^2|^2 + 2|X y^U_{11} y^U_{21} + Z y^W_{12} y^W_{22}|^2 + |X (y^U_{21})^2 + Z (y^W_{22})^2|^2 \\
    & + 2|X y^U_{11} y^U_{31} + Z y^W_{12} y^W_{32}|^2 + |X (y^U_{31})^2 + Z (y^W_{32})^2|^2 + 2 |X y^U_{21} y^U_{31} + Z y^W_{22} y^W_{32} |^2,
  \end{split}
\end{align}
where the element notation for Eqs.(\ref{eq:32transformedYukawa1}-\ref{eq:32transformedYukawa2}) is used, e.g., $y^W_{i2}=[y^W_{\nu2}(t_1)]_i$.
Importantly, the invariant $I_0$ can be rewritten as the complex product $I_0=\mathcal{I}\times \mathcal{I}^\ast$.
Then $\mathcal{I}$ is written as the matrix elements from $[\overset{(1)}{m_{eff}}(t_1)]_{ij}=m_{ij}$,
\begin{equation}
  \mathcal{I} \equiv m_{33}(m_{11}m_{22}-m_{21}m_{12})
    +m_{13}(m_{32}m_{21}-m_{31}m_{22})
    +m_{32}(m_{31}m_{12}-m_{32}m_{11}).
    \label{eq:brokenInvariant}
\end{equation}
After substitution of the matrix elements from Eq.(\ref{eq:TransformedMass32}) the terms of $\mathcal{I}$ are,
\begin{gather}
  m_{33}(m_{11}m_{22}-m_{21}m_{12}) = \frac{v^3}{8}XZ\left(y^W_{12}y^U_{21}-y^U_{11}y^W_{22}\right)^2\left(X(y^U_{31})^2+Z(y^W_{32})^2\right),
  \label{eq:32elementsI01}
\\
  m_{13}(m_{32}m_{21}-m_{31}m_{22}) = -\frac{v^3}{8}XZ \left(y^W_{12}y^U_{21}-y^U_{11}y^W_{22}\right)\left(y^U_{21}y^W_{32}-y^W_{22}y^U_{31}\right)\left(Xy^U_{11} y^U_{31}+Zy^W_{12}y^W_{32}\right),
  \label{eq:32elementsI02}
\\
  m_{32}(m_{31}m_{12}-m_{32}m_{11}) = -\frac{v^3}{8}XZ \left(y^W_{12}y^U_{21}-y^U_{11}y^W_{22}\right)\left(y^W_{12}y^U_{31}-y^U_{11}y^W_{32}\right)\left(Xy^U_{21} y^U_{31}+Zy^W_{22}y^W_{32}\right)
  \label{eq:32elementsI03}.
\end{gather}
The addition of the last two Eqs.(\ref{eq:32elementsI02}-\ref{eq:32elementsI03}) exactly cancel the first Eq.(\ref{eq:32elementsI01}), which leads to $\mathcal{I}=0$ that proves the $\rank[\overset{(1)}{m_{eff}}(t_1)]=2$.
In other words, the lightest mass eigenvalue is zero $\overset{(1)}{m_{lightest}}(t_1)=0$ as typical for 3-2 Seesaw models.
This result does not depend on the details of how the kernel matrices affect the Yukawa vectors.
Then from consideration of the invariants we conclude that,
\begin{equation}
 \rank\left[\overset{(1)}{\kappa}(t_1)\right]=\rank\left[\overset{(1)}{m_{eff}}(t_1)\right]=2.
\end{equation}
This result for the 3-2 model is the one-loop renormalization group equations and tree-level matching.
However, the result has been shown to not hold for one-loop matching in supersymmetry \cite{Zhou:2021bqs} and two-loop renormalization group equation for the Weinberg operator coefficient $\kappa$ \cite{Davidson:2006tg}.
Inclusion of all two-loop renormalization group equations \cite{Ibarra:2018dib,Ibarra:2020eia} and complete one-loop level matching in the Standard Model could be interesting.
Here we use the one-loop equations and tree-level matching to prepare, for section \ref{sec:33Seesaw}, on how a rank degenerate matrix remains that way even after renormalization group effects are considered.

\section{3-3 Type-I Seesaw Model Renormalization Group Running}
\subsection{Effective mass matrix with generic renormalization group kernel solutions}\label{sec:33Seesaw}
\begin{figure}[htb]
  \centering
  \includegraphics[width=125mm]{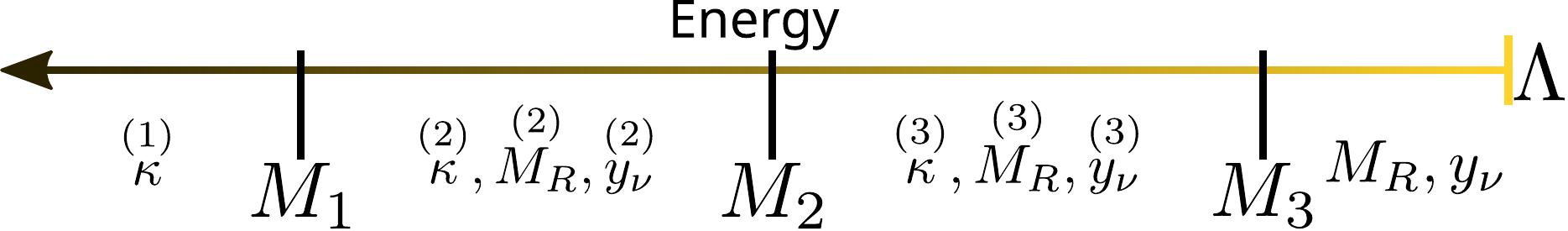}
  \caption{\label{Fig:3EnergyScales}
  Illustration of the different energy regions being considered for the 3-3 model.
  Similar to Fig \ref{Fig:2EnergyScales}, the highest energy region of the full theory occurs between $M_3$ and $\Lambda$ where no $\kappa$ term exists.
  The renormalization energy scale decreases from right to the left.
  The matrices $\kappa(\mu)$, $y_\nu(\mu)$, and $M_R(\mu)$ are scale dependent based on the $\beta$-functions $\beta_\kappa$, $\beta_{y_\nu}$, and $\beta_{M_R}$ defined by the corresponding renormalization group equations.}
\end{figure}

In this section we consider the next simplest version of the Type-I seesaw model.
The Lagrangian of Eq.(\ref{Eq:Lagrangian}) remains the same, but now three right-handed neutrinos are included $N_1$, $N_2$, and $N_3$.
We assume the masses of the right-handed neutrinos are hierarchical with $M_3 > M_2 > M_1$ as illustrated in Fig \ref{Fig:3EnergyScales}.
This model is described by a full theory at the energy scale of $\mu>M_3$ up to some cutoff $\Lambda$.
We are interested in the case when the effective mass matrix is rank degenerate between the energy scale of $M_3< \mu \le \Lambda$.
Meaning we create the condition of,
\begin{align}
   m_{eff}&=-\frac{v^2}{2}y_{\nu}\frac{1}{M_{R}}y^T_{\nu}, \label{eq:meff} \\
   \rank\left[m_{eff}\right]&=2,
\end{align}
by assuming the Yukawa matrix is also rank degenerate,
\begin{align}
   y_{\nu i j }&=\begin{pmatrix} y_{\nu i1} & y_{\nu i 2} & y_{\nu i 3} \end{pmatrix}, \\
 y_{\nu 3}&=ay_{\nu 1}+by_{\nu 2}, \label{eq:33HighEnergyYukawa}\\
   \rank\left[y_{\nu}\right]&=2 .
\end{align}
In Eq.(\ref{eq:33HighEnergyYukawa}), $a$ and $b$ are complex-numbers of arbitrary values.
The effective mass matrix of Eq.(\ref{eq:meff}) can be rewritten in a similar form of the 3-2 model by using our assumption of Eq.(\ref{eq:33HighEnergyYukawa});
\begin{equation}
  m_{eff}=-\frac{v^2}{2}
    \begin{pmatrix} y_{\nu 1} & y_{\nu 2} \end{pmatrix} 
    \begin{pmatrix} \frac{1}{M_1}+\frac{a^2}{M_3} & \frac{a b}{M_3} \\
      \frac{a b}{M_3}  & \frac{1}{M_2}+ \frac{b^2}{M_3} \end{pmatrix} 
    \begin{pmatrix} y^T_{\nu 1} \\ y^T_{\nu 2}  \end{pmatrix},
  \label{Eq:33ModelLinear}
\end{equation}
where we take the diagonal basis of the right-handed neutrino mass matrix $M_R$.
Our condition of $\rank\left[m_{eff}\right]=2$ holds for the entire energy region of $M_3 < \mu \le \Lambda$.
We prove this by considering the solutions of the renormalization group equations of Eqs.(\ref{eq:Kernelyukawa}-\ref{eq:KernelRightMass}),
\begin{gather}
  Y_{\nu i}(\mu) = U^\prime (\mu)y_{\nu 1}\left\{[K^{-1}(\mu)]_{i1}+a[K^{-1}(\mu)]_{i3}\right\}
  +U^\prime (\mu)y_{\nu 2}\left\{[K^{-1}(\mu)]_{i2}+b[K^{-1}(\mu)]_{i3}\right\}, \\
  Y_{\nu 3}(\mu)=A(\mu)Y_{\nu 1}(\mu)+B(\mu)Y_{\nu 2}(\mu),
\end{gather}
where, $i=1,2$. $U^\prime(\mu)$ and $K(\mu)$  denote  $3 \times 3$ evolution matrices for Yukawa vectors and Majorana mass matrix respectively, for the region $M_3 < \mu \le \Lambda$.
$A(\mu)$ and $B(\mu)$ are constructed from elements of $K^{-1}(\mu)$, $a$, and $b$.
This results in Eq.(\ref{Eq:33ModelLinear}) becoming,
\begin{equation}
  m_{eff}(\mu)=-\frac{v^2}{2}
  \begin{pmatrix} Y_{\nu 1}(\mu) & Y_{\nu 2}(\mu) \end{pmatrix} 
  \begin{pmatrix} \frac{1}{M_1}+\frac{A^2(\mu)}{M_3} & \frac{A(\mu) B(\mu)}{M_3} \\
    \frac{A(\mu) B(\mu)}{M_3}  & \frac{1}{M_2}+ \frac{B^2(\mu)}{M_3} \end{pmatrix} 
  \begin{pmatrix} Y^T_{\nu 1}(\mu) \\ Y^T_{\nu 2}(\mu)  \end{pmatrix},
\end{equation}
which maintains the condition of $\rank\left[m_{eff}\right]=2$.

The difference between Eq.(\ref{Eq:33ModelLinear}) of the 3-3 model and the 3-2 model is how the effective mass matrix is rank degenerate.
In the 3-2 model, the rank degeneracy occurs because the third right-handed neutrino is absent.
Whereas, in the 3-3 model we impose an additional condition on the Yukawa coupling $y_{\nu 3}$ between the heavy and light neutrino Eq.(\ref{eq:33HighEnergyYukawa});
such that $y_{\nu 3}$ is described by a linear combination of the other two Yukawa couplings, leading to a rank degenerate effective mass matrix.
As we saw in the 3-2 model in section \ref{sec:32Seesaw}, the running of the renormalization group equations can not create a rank three effective mass matrix.
Therefore, the lightest neutrino mass is zero even if we take into account the renormalization group effects.
Thus, considering Eq.(\ref{Eq:33ModelLinear}) we are interested if the 3-3 model has the rank degenerate effective mass matrix at low energy similar to the 3-2 model.

To begin, we study the renormalization group effects with the same processes as the 3-2 model. 
We start at the high energy scale of $\mu = M_3$ or $t = 0$ where,
\begin{equation}
 t \equiv \log \frac{M_3}{\mu}, \quad t_2 \equiv \log \frac{M_3}{M_2}, \quad t_1 \equiv \log \frac{M_3}{M_1}.
\end{equation}
Without loss of generality, one can start with the diagonal basis for the right-handed Majorana neutrinos, based on the discussion after Eq.(\ref{Eq:33ModelLinear}).
To go to lower energies we must integrate out the heaviest neutrino $N_3$, which generates the effective coefficient $\overset{(3)}{\kappa}(t)$.
Matching of the effective mass matrix across $M_3$ provides initial conditions of the renormalization group equations for $\overset{(3)}{\kappa}(t)$, $\overset{(3)}{M_R}(t)$, and $\overset{(3)}{y_{\nu}}(t)$.
Resulting in equations similar to the 3-2 model,
\begin{subequations}
  \begin{align}
   \overset{(3)}{m_{eff}}(0) &= m_{eff}(0) , \\
    -\frac{v^2}{4}\overset{(3)}{\kappa}(0)
    -\frac{v^2}{2}\overset{(3)}{y_{\nu}}(0)\left(\overset{(3)}{M_R}(0)\right)^{-1}\left(\overset{(3)}{y_{\nu}}(0)\right)^T
&=-\frac{v^2}{2}y_{\nu}(0)\frac{1}{M_{R}(0)}y^T_{\nu}(0).
  \end{align} \label{eq:33initialMatching}
\end{subequations}
Using Eq.(\ref{eq:33initialMatching}), one can determine $\overset{(3)}{\kappa}(0)$ as,
\begin{equation}
  \overset{(3)}{\kappa}(0)=2 y_{\nu 3}(0)\frac{1}{M_3(0)}y^T_{\nu 3}(0),
\end{equation}
where $y_{\nu 3}(0)=ay_{\nu 1}(0)+by_{\nu 2}(0)$ from Eq.(\ref{eq:33HighEnergyYukawa}).

Next, we integrate down to the mass of $M_{2}$ by solving the renormalization group equations in the same way as Eqs.(\ref{eq:KappaRGESolution}-\ref{eq:RightMassRGESolution}).
In contrast to the 3-2 model, an additional energy scale remains between the masses of $M_2$ and $M_1$ before we reach the lowest scale of $\mu = M_1$.
To account for this additional region we first must integrate out the right-handed neutrino $N_2$.
The mass of $N_2$ is not the same as $M_2(0)$ due to renormalization group corrections from the running of $\overset{(3)}{M_R}(0)$ generating off-diagonal components in $\overset{(3)}{M_R}(t_2)$.
So we diagonalize the matrix $\overset{(3)}{M_R}(t_2)$ with a Takagi decomposition 
\cite{Takagi:1925, Autonne:1915, Hahn:2006hr},
\begin{equation}
 \overset{(3)}{\mathbf{M}_R}(t_2)
 = V^T\overset{(3)}{M_R}(t_2)V ,
 \label{eq:33diagTakagi}
\end{equation}
where $V$ is a $2 \times 2$ unitary matrix.
The diagonal matrix $ \overset{(3)}{\mathbf{M}_R}(t_2)$ is $ \overset{(3)}{\mathbf{M}_{R\,ij}}(t_2)=\delta_{ij}  \overset{(3)}{\mathbf{M}_{i}}(t_2)$. 
In the diagonal basis $\overset{(3)}{\mathbf{M}_R}(t_2)$,  the Yukawa matrix $\overset{(3)}{\mathbf{y}_{\nu}}(t_2)$  is given by the $3 \times 2$ matrix,
\begin{equation}
  \overset{(3)}{\mathbf{y}_{\nu}}(t_2)= \overset{(3)}{{y}_{\nu}}(t_2) V=
  \begin{pmatrix} \overset{(3)}{\mathbf{y}_{\nu 1}}(t_2) &
   \overset{(3)}{\mathbf{y}_{\nu 2}}(t_2)
  \end{pmatrix}.
  \label{eq:yukawa23}
\end{equation}
Then we integrate out the right-handed neutrino $N_2$ in the diagonal basis and the effective mass matrix is matched across the energy of $M_2$ in a manner similar to Eq.(\ref{eq:33initialMatching}),
\begin{align} \label{eq:33secondMatching}
  \overset{(2)}{m_{eff}}(t_2) &= \overset{(3)}{m_{eff}}(t_2),
\\
   -\frac{v^2}{4}\overset{(2)}{\kappa}(t_2)
    -\frac{v^2}{2}\overset{(2)}{y_{\nu}}(t_2)
    \left(\overset{(2)}{M_R}(t_2)\right)^{-1}
    \overset{(2)}{y_{\nu}}(t_2)^T &=-\frac{v^2}{4}\overset{(3)}{\kappa}(t_2)
    -\frac{v^2}{2}\overset{(3)}{\mathbf{y}_{\nu}}(t_2)
    \left(\overset{(3)}{\mathbf{M}_R}(t_2)\right)^{-1}
    \overset{(3)}{\mathbf{y}_{\nu}}(t_2)^T,
\end{align}
where $\overset{(2)}{M_R}(t_2)=\overset{(3)}{\mathbf{M}_{1}}(t_2) $ and $\overset{(2)}{y_{\nu}}(t_2)=\overset{(3)}{\mathbf{y}_{\nu 1}}(t_2)$.
In the left-handed side of Eq.(\ref{eq:33secondMatching}), a new Wilson coefficient $\overset{(2)}{\kappa}(t_2)$ is generated and is determined as,
\begin{equation}
  \overset{(2)}{\kappa}(t_2) =2~\overset{(3)}{\mathbf{y}_{\nu 2}}(t_2)\left(\overset{(3)}{\mathbf{M}_{2}}(t_2)\right)^{-1}\overset{(3)}{\mathbf{y}_{\nu 2}}(t_2)^{T}+\overset{(3)}{\kappa}(t_2).
\label{eq:kappat2}
\end{equation}
Lastly, we solve the renormalization group equations for $\overset{(2)}{\kappa}(t)$, $\overset{(2)}{M_R}(t)$, and $\overset{(2)}{y_{\nu}}(t)$ between the energies of $N_2$ and $N_1$.
This results in an effective mass matrix of,
\begin{equation}
  \overset{(1)}{m_{eff}}(t_1)=-\frac{v^2}{4}\overset{(2)}{\kappa}(t_1)
  -\frac{v^2}{2}\overset{(2)}{y_{\nu}}(t_1)
  \left(\overset{(2)}{M_R}(t_1)\right)^{-1}
  \overset{(2)}{y_{\nu}}(t_1)^T.
\label{eq:meff1}
\end{equation}
After which, $N_1$ would need to be integrated out to continue going down to the low energy experimental region $\mu < M_1$.
We are not interested in what occurs for $\mu < M_1$ as it will not influence the rank of the effective mass matrix $\overset{(1)}{m_{eff}}(t)$.

To investigate the rank of the effective mass matrix, we rewrite Eq.(\ref{eq:meff1}) by substituting the kernel functions for the region $[t_1,t_2)$ explicitly.
From Eq.(\ref{eq:YukawaRGESolution}) one obtains,
\begin{equation}
  \overset{(2)}{y_{\nu}}(t_1)=e^{-\frac{1}{16\pi^2}\int_{t_2}^{t_1}\overset{(2)}{\alpha_{y_\nu}}(s) ds}\overset{(2)}{U}(t_1, t_2)\overset{(2)}{y_{\nu}}(t_2).
 \label{eq:y2t2}
\end{equation}
Then we explicitly write the kernel function which relates $\overset{(2)}{\kappa}(t_1)$ to $\overset{(2)}{\kappa}(t_2)$,
\begin{align}
  \overset{(2)}{\kappa}(t_1)={}&e^{-\frac{1}{16\pi^2}\int_{t_2}^{t_1}\overset{(2)}{\alpha_\kappa}(s) ds}\overset{(2)}{W}(t_1-t_2)\overset{(2)}{\kappa}(t_2)\overset{(2)}{W}(t_1-t_2)^T,
\\
  \begin{split}
    ={}&\frac{4q}{v^2}\overset{(2)}{W}(t_1-t_2)\overset{(3)}{\mathbf{y}_{\nu 2}}(t_2)\overset{(3)}{\mathbf{y}_{\nu 2}}(t_2)^{T} \overset{(2)}{W}(t_1-t_2)^T
  \\
    &+\frac{4x}{v^2}\overset{(2)}{W}(t_1-t_2)\overset{(3)}{W}(t_2)\overset{(3)}{y_\nu}(0) 
    \begin{pmatrix} a^2 & a b \\ a b & b^2 \end{pmatrix} 
    \overset{(3)}{y_\nu}(0)^T \overset{(3)}{W}(t_2)^T\overset{(2)}{W}(t_1-t_2)^T.
    \label{eq:kappat1}
  \end{split}
\end{align}
We have gathered the scaling terms into the factors,
\begin{gather}
   q \equiv  \frac{v^2}{2\overset{(3)}{\mathbf{M}_{2}}(t_2)}e^{-\frac{1}{16\pi^2}\int_{t_2}^{t_1}\overset{(2)}{\alpha_\kappa}(s) ds},
\\
  x \equiv \frac{v^2}{2 M_3(0)}e^{-\frac{1}{16\pi^2}\int_{t_2}^{t_1}\overset{(2)}{\alpha_\kappa}(s) ds}e^{-\frac{1}{16\pi^2}\int_{0}^{t_2}\overset{(1)}{\alpha_\kappa}(s) ds},
\\
  z \equiv \frac{v^2}{2 \overset{(2)}{M_R}(t_1)}e^{-\frac{1}{8\pi^2}\int_{t_2}^{t_1}\overset{(2)}{\alpha_{y_\nu}}(s) ds},
\end{gather}
because they will not modify the rank of the effective mass matrix.
After the substitution of  Eq.(\ref{eq:y2t2}) and Eq.(\ref{eq:kappat1}) into Eq.(\ref{eq:meff1}), $\overset{(1)}{m_{eff}}(t_1)$ is rewritten as,
\begin{equation}
  \begin{split}
    \overset{(1)}{m_{eff}}(t_1)={}&-q\overset{(2)}{W}(t_1-t_2)\overset{(3)}{\mathbf{y}_{\nu 2}}(t_2)\overset{(3)}{\mathbf{y}_{\nu 2}}(t_2)^{T} \overset{(2)}{W}(t_1-t_2)^T
  \\
    &-x\overset{(2)}{W}(t_1-t_2)\overset{(3)}{W}(t_2)\overset{(3)}{y_\nu}(0) 
    \begin{pmatrix} a^2 & a b \\ a b & b^2 \end{pmatrix} 
    \overset{(3)}{y_\nu}(0)^T \overset{(3)}{W}(t_2)^T\overset{(2)}{W}(t_1-t_2)^T
  \\
    &-z\overset{(2)}{U}(t_1-t_2)\overset{(2)}{y_{\nu}}(t_2)
    \overset{(2)}{y_{\nu}}(t_2)^T \overset{(2)}{U}(t_1-t_2)^T.
  \end{split}
  \label{eq:meff1Interp}
\end{equation}

With the effective mass matrix given in Eq.(\ref{eq:meff1Interp}), the mass squared eigenvalues of the Hermite matrix $\overset{(1)}{\mathcal{M}}=\overset{(1)}{m_{eff}}(t_1) [\overset{(1)}{m_{eff}}(t_1)]^\dagger$ does not clearly have a zero eigenvalue.
To prove this we evaluate the three invariants in Eqs.(\ref{eq:invariant2}-\ref{eq:invariant0}).
The first two invariants $I_1\neq0$ and $I_2\neq0$ are non-zero in a manner similar to the Eqs.(\ref{eq:32I1result}-\ref{eq:32I2result}) of the 3-2 model.
Recall, the invariant $I_0$ can be rewritten as the complex product $I_0=\mathcal{I}\times \mathcal{I}^\ast$.
Then $\mathcal{I}$ is written as the matrix elements from $[\overset{(1)}{m_{eff}}(t_1)]_{ij}=m_{ij}$ as in Eq.(\ref{eq:brokenInvariant}),
\begin{gather}
  \begin{split}
    m_{33}(m_{11}m_{22}-m_{21}m_{12}) =&{}\left[-q(y^W_{32})^2-x(ay^{WW}_{31}+by^{WW}_{32})^2-z(y^U_{11})^2\right]
  \\
    &\times\left\{-\left[qy^W_{12}y^W_{22}+x(ay^{WW}_{11}+by^{WW}_{12})(ay^{WW}_{21}+by^{WW}_{22})+zy^U_{11}y^U_{21}\right]^2\right.
  \\
    &\qquad+[q(y^{W}_{12})^2+x(ay^{WW}_{11}+by^{WW}_{12})^2+(y^U_{11})^2]
  \\
    &\Bigl.\qquad\times[q(y^W_{22})^2+x(ay^{WW}_{21}+by^{WW}_{22})^2+z(y^U_{21})^2]\Bigr\},
  \label{eq:33elementsI01}
  \end{split}
\\ 
  \begin{split}
    m_{13}(m_{32}m_{21}-m_{31}m_{22}) ={}& \left[-qy^W_{12}y^W_{32}-x(ay^{WW}_{11}+by^{WW}_{12})(ay^{WW}_{31}+by^{WW}_{32})-zy^U_{11}y^U_{31}\right]
  \\
    &\,\left\{-\left[(q(y^W_{22})^2+x(ay^{WW}_{21}+by^{WW}_{12})^2+z(y^U_{21})^2)\right.\right.
  \\
    &\left.\qquad\times(qy^W_{12}y^W_{32}+x(ay^{WW}_{11}+by^{WW}_{12})(ay^{WW}_{31}+by^{WW}_{32})+zy^U_{11}y^U_{31})\right]
  \\
    &\quad+(qy^W_{12}y^W_{22}+x(ay^{WW}_{11}+by^{WW}_{12})(ay^{WW}_{21}+by^{WW}_{22})+zy^U_{11}y^U_{21})
  \\
    &\left.\quad\times(qy^W_{22}y^W_{32}+x(ay^{WW}_{21}+by^{WW}_{22})(ay^{WW}_{31}+by^{WW}_{32})+zy^U_{21}y^U_{31})\right\},
    \label{eq:33elementsI02}
  \end{split}
\\ 
  \begin{split}
    m_{32}(m_{31}m_{12}-m_{32}m_{11}) ={}& \left[-qy^W_{22}y^W_{32}-x(ay^{WW}_{21}+by^{WW}_{22})(ay^{WW}_{31}+by^{WW}_{32})-zy^U_{21}y^U_{31}\right]
  \\
    &\,\left\{-\left[(q(y^W_{12})^2+x(ay^{WW}_{11}+by^{WW}_{12})^2+z(y^U_{11})^2)\right.\right.
  \\
    &\left.\qquad\times(qy^W_{22}y^W_{32}+x(ay^{WW}_{21}+by^{WW}_{22})(ay^{WW}_{31}+by^{WW}_{32})+zy^U_{21}y^U_{31})\right]
  \\
    &\quad+(qy^W_{12}y^W_{22}+x(ay^{WW}_{11}+by^{WW}_{12})(ay^{WW}_{21}+by^{WW}_{22})+zy^U_{11}y^U_{21})
  \\
    &\left.\quad\times(qy^W_{12}y^W_{32}+x(ay^{WW}_{11}+by^{WW}_{12})(ay^{WW}_{31}+by^{WW}_{32})+zy^U_{11}y^U_{31})\right\}.
    \label{eq:33elementsI03}
  \end{split}
\end{gather}
We define the following Yukawa vectors from Eq.(\ref{eq:meff1Interp}),
\begin{align}
  y^W_{\nu 2}(t_1)&=\overset{(2)}{W}(t_1-t_2)\overset{(3)}{\mathbf{y}_{\nu 2}}(t_2), \\
y^{WW}_{\nu j}(t_1)&=\overset{(2)}{W}(t_1-t_2)\overset{(3)}{W}(t_2)\overset{(3)}{y_{\nu j}}(0), \qquad \text{where }j=1,2 , \\
y^{U}_{\nu 1}(t_1)&=\overset{(2)}{U}(t_1-t_2)\overset{(2)}{y_{\nu}}(t_2).
\end{align}
The matrix elements in the determinant of $m_{ij}$ in Eqs.(\ref{eq:33elementsI01}-\ref{eq:33elementsI03}), are identified as;
\begin{align}
  y^{WW}_{i j}=[y^{WW}_{\nu j}]_i, && y^W_{i 2}=[y^W_{\nu 2}]_i,&& y^U_{i 1}=[y^U_{\nu 1}]_i,&& \text{where }i=1,2,3.
\end{align}
Unlike the results for the 3-2 seesaw model, the terms of Eq.(\ref{eq:33elementsI01}) and Eq.(\ref{eq:33elementsI02}) do not cancel with Eq.(\ref{eq:33elementsI03}) under addition, leading us to no definite conclusions.
This is because of the three different kernel matrix transformations acting on the Yukawa vectors are not guarantied to act similarly.
Thus, for a definite conclusion on the rank of the effective mass matrix we must investigate the details of the kernel matrices.

\subsection{The rank of the effective mass matrix with specific renormalization group kernel solutions}\label{sec:33SeesawRank}
In this section, we examine the rank of the effective Majorana mass matrix Eq.(\ref{eq:meff1Interp}) at low energy.
Since the rank depends on the kernel functions, one must specify them.
We examine the rank for the case that all the renormalization group evolution is governed by the Yukawa couplings for neutrinos\footnote{Any effects of the charged lepton Yukawa couplings will change our proof and may alter the present result.}.
In appendix \ref{sec:reverseKappaproof} we prove the relation of Eq.(\ref{eq:Eq.b2}), which we will use now;
\begin{equation}
  \begin{split}
    \overset{(3)}{W}(t_2) \overset{(3)}{y_{\nu}}(0)& = T \exp\left[-\frac{1}{16\pi^2}\int_{0}^{t_2}(\overset{(3)}{y_{\nu}}(s)\overset{(3)}{y_{\nu}}(s)^{\dagger}) ds \right] \overset{(3)}{y_{\nu}}(0)
\\
    &=\overset{(3)}{y_{\nu}}(t_2) T \exp\left[\frac{1}{32\pi^2}\int_{0}^{t_2}
  (\overset{(3)}{y_{\nu}}(s)^\dagger \overset{(3)}{y_{\nu}}(s)) ds \right] e^{\frac{1}{16\pi^2}\int^{t_2}_0 \overset{(3)}{\alpha_{y_\nu}}(s) ds}
\\
    &=\overset{(3)}{y_{\nu}}(t_2)\overset{(3)}{W_R}(t_2) e^{\frac{1}{16\pi^2}\int^{t_2}_0 \overset{(3)}{\alpha_{y_\nu}}(s) ds},
  \end{split}
 \label{eq:Eq.b1}
\end{equation}
where we have defined,
\begin{equation}
  \overset{(3)}{W_R}(t_2)=T \exp\left[\frac{1}{32\pi^2}\int_{0}^{t_2}
  (\overset{(3)}{y_{\nu}}(s)^\dagger \overset{(3)}{y_{\nu}}(s)) ds \right].
  \label{eq:Eq.WR}
\end{equation}
Using the reverse Kernel relation of Eq.(\ref{eq:Eq.b1}), we rewrite the middle term of Eq.(\ref{eq:meff1Interp}) as follows;
\begin{equation}
  \begin{split}
    \overset{(1)}{m_{eff}}(t_1)={}&-q\overset{(2)}{W}(t_1-t_2)\overset{(3)}{\mathbf{y}_{\nu 2}}(t_2)\overset{(3)}{\mathbf{y}_{\nu 2}}(t_2)^{T} \overset{(2)}{W}(t_1-t_2)^T
\\
    &-x^{\prime}\overset{(2)}{W}(t_1-t_2)\overset{(3)}{y_\nu}(t_2) \overset{(3)}{W_R}(t_2)
    \begin{pmatrix} a^2 & a b \\ a b & b^2 \end{pmatrix} 
    (\overset{(3)}{W_R}(t_2))^T  \overset{(3)}{y_\nu}(t_2)^T \overset{(2)}{W}(t_1-t_2)^T
\\
    &-z\overset{(2)}{U}(t_1-t_2)\overset{(2)}{y_{\nu}}(t_2)\overset{(2)}{y_{\nu}}(t_2)^T \overset{(2)}{U}(t_1-t_2)^T,
    \label{eq:meff1Interp2}
  \end{split}
\end{equation}
where $x^{\prime}=x e^{\frac{1}{8\pi^2}\int^{t_2}_0 \overset{(3)}{\alpha_{y_\nu}}(s) ds}$.
Next we continue focusing on the second line of Eq.(\ref{eq:meff1Interp2}) where $\overset{(3)}{y_\nu}(t_2)$  is a 3 by 2 matrix.
We rewrite $\overset{(3)}{y_\nu}(t_2)$ following Eq.(\ref{eq:yukawa23}) that comes from diagonalizing the right-handed mass matrix at $t_2$,
\begin{equation}
  \overset{(3)}{{y}_{\nu}}(t_2)=\overset{(3)}{\mathbf{y}_{\nu}}(t_2) V^{-1}= 
\begin{pmatrix} \overset{(3)}{\mathbf{y}_{\nu 1}}(t_2) & \overset{(3)}{\mathbf{y}_{\nu 2}}(t_2)  \end{pmatrix} V^{-1} 
=\begin{pmatrix} \overset{(2)}{y_{\nu}}(t_2) & \overset{(3)}{\mathbf{y}_{\nu 2}}(t_2)  \end{pmatrix} V^{-1}
\end{equation}
where we use the relation $\overset{(2)}{{y}_{\nu}}(t_2)=\overset{(3)}{\mathbf{y}_{\nu 1}}(t_2)$.
Then one can rewrite Eq.(\ref{eq:meff1Interp2}) into the following form,
\begin{equation}
  \begin{split}
    \overset{(1)}{m_{eff}}(t_1)={}&-q\overset{(2)}{W}(t_1-t_2)\overset{(3)}{\mathbf{y}_{\nu 2}}(t_2)\overset{(3)}{\mathbf{y}_{\nu 2}}(t_2)^{T} \overset{(2)}{W}(t_1-t_2)^T
\\
    &-x^\prime \overset{(2)}{W}(t_1-t_2)
    \begin{pmatrix} \overset{(2)}{y_\nu}(t_2) & \overset{(3)}{\mathbf{y}_{\nu 2}}(t_2)  \end{pmatrix} A(t_2) \begin{pmatrix} \overset{(2)}{y_{\nu}}(t_2)^T \\ \overset{(3)}{\mathbf{y}_{\nu 2}}(t_2)^T \end{pmatrix}\overset{(2)}{W}(t_1-t_2)^T
\\
    &-z\overset{(2)}{U}(t_1-t_2)\overset{(2)}{y_{\nu}}(t_2)\overset{(2)}{y_{\nu}}(t_2)^T \overset{(2)}{U}(t_1-t_2)^T,
    \label{eq:mefft1}
  \end{split}
\end{equation}
where $A$ is a two by two matrix defined as,
\begin{equation}
  A(t_2)\equiv V^{-1}  \overset{(3)}{W}_R(t_2)\begin{pmatrix} a^2 & a b \\ a b & b^2 \end{pmatrix}
(V^{-1} \overset{(3)}{W}_R(t_2) )^T.
\end{equation}
Similar to Eqs.(\ref{eq:Eq.b1}, \ref{eq:Eq.WR}, \ref{eq:Eq.b2}), one can show the following relation,
\begin{equation}
  \overset{(2)}{W}(t_1-t_2) \overset{(2)}{y_\nu}(t_2) = \overset{(2)}{y_{\nu}}(t_1) \overset{(2)}{W_R}(t_1-t_2)
e^{\frac{1}{16\pi^2}\int^{t_1}_{t_2} \overset{(2)}{\alpha_{y_\nu}}(s) ds},
  \label{eq:Eq.bt1t2}
\end{equation}
where $\overset{(2)}{W_R}(t_1-t_2)$ is defined by,
\begin{equation}
  \overset{(2)}{W_R}(t_1-t_2)
  = T \exp\left[\frac{1}{32\pi^2}\int_{t_2}^{t_1}(\overset{(2)}{y_{\nu}}(s)^\dagger \overset{(2)}{y_{\nu}}(s)) ds \right].
\end{equation}
Further one defines,
\begin{equation}
  \overset{W}{\mathbf{y}_{\nu 2}}(t_1) \equiv \overset{(2)}{W}(t_1-t_2) \overset{(3)}{\mathbf{y}_{\nu 2}}(t_2).
  \label{eq:Eq.yw}
\end{equation}
Combining Eq.(\ref{eq:Eq.bt1t2}) with Eq.(\ref{eq:Eq.yw}), one obtains the following  relation that can rewrite the second line of Eq.(\ref{eq:mefft1}).
\begin{equation}
  \overset{(2)}{W}(t_1-t_2)
  \begin{pmatrix} \overset{(2)}{y_\nu}(t_2) & \overset{(3)}{\mathbf{y}_{\nu 2}}(t_2)  \end{pmatrix} 
  =\begin{pmatrix} \overset{(2)}{y_\nu}(t_1) & \overset{W}{\mathbf{y}_{\nu 2}}(t_1)  \end{pmatrix}
  \begin{pmatrix} \overset{(2)}{W_R^{\prime}}(t_1-t_2)& 0 \\ 0 & 1 \end{pmatrix},
\end{equation}
where $\overset{(2)}{W_R^{\prime}}(t_1-t_2)=\overset{(2)}{W_R}(t_1-t_2) e^{\frac{1}{16\pi^2}\int^{t_1}_{t_2}  \overset{(2)}{\alpha_{y_\nu}}(s) ds}$.
Then $\overset{(1)}{m_{eff}}(t_1)$ has the following form,
\begin{equation}
  \begin{split}
    \overset{(1)}{m_{eff}}(t_1)={}&-q\,\overset{W}{\mathbf{y}_{\nu 2}}(t_1)\overset{W}{\mathbf{y}_{\nu 2}}(t_1)^{T}
  \\
    &-x^\prime \begin{pmatrix} \overset{(2)}{y_\nu}(t_1) & \overset{W}{\mathbf{y}_{\nu 2}}(t_1)  \end{pmatrix}
    \begin{pmatrix} \overset{(2)}{W^{\prime}_R}(t_1-t_2) & 0 \\ 0 & 1 \end{pmatrix}
    A(t_2)
    \begin{pmatrix} \overset{(2)}{W^{\prime}_R}(t_1-t_2) & 0 \\ 0 & 1 \end{pmatrix}
    \begin{pmatrix} \overset{(2)}{y_{\nu}}(t_1)^T \\ \overset{W}{\mathbf{y}_{\nu 2}}(t_1)^T \end{pmatrix}
  \\
    &-\frac{v^2}{2 \overset{(2)}{M_R}(t_1)}\overset{(2)}{y_{\nu}}(t_1)\overset{(2)}{y_{\nu}}(t_1)^T
  \end{split}
  \label{eq:33meffTwovectorform}
\end{equation}
Lastly, we can see only two Yukawa vectors remain in Eq.(\ref{eq:33meffTwovectorform}), $\overset{(2)}{y_\nu}(t_1)$ and $\overset{W}{\mathbf{y}_{\nu 2}}(t_1)$.
This proves that the third Yukawa vector is constructed as a linear combination of the other two.
Next, we rewrite $\overset{(1)}{m_{eff}}(t_1)$ to highlight this result,
\begin{equation}
  \overset{(1)}{m_{eff}}(t_1)=
 \begin{pmatrix}
  \overset{(2)}{y_\nu}(t_1) & \overset{W}{\mathbf{y}_{\nu 2}}(t_1) 
 \end{pmatrix}
 \begin{pmatrix}
  \alpha(t_1,t_2) & \beta(t_1,t_2)
  \\
  \beta(t_1,t_2) & \gamma(t_1,t_2)
 \end{pmatrix}
 \begin{pmatrix} 
  \overset{(2)}{y_{\nu}}(t_1)^T
  \\
  \overset{W}{\mathbf{y}_{\nu 2 }}(t_1)^T
 \end{pmatrix}.
\end{equation}

As a result, for the simplest kernel functions the two Yukawa vectors lead us to conclude that the effective mass matrix is rank degenerate,
\begin{equation}
 \rank \left[\overset{(1)}{m_{eff}}(t_1) \right]  \le 2 .
\end{equation}
This means that the determinant of the matrix is zero, thus the lightest active neutrino is massless.
The implication is in the full theory energy region the lightest active neutrino is massless, and even after evolution of the renormalization group equations to below the scale of $M_3$ that same neutrino is massless.


\section{Conclusion}\label{sec:Conclusion}
We studied the effect of the renormalization group equations on the rank of the effective mass matrix for active Majorana neutrinos. 
The Type-I seesaw models are examined with two and three right-handed neutrinos with hierarchical masses.
The right-handed neutrinos are integrated one by one at their thresholds, and it generates the Weinberg operator.
For the 3-2 model, with two right-handed Majorana neutrinos, the rank of the effective mass matrix is two at high energy scales;
and the renormalization group equations do not change the rank at the energy scale lower than the lightest right-handed neutrino $N_1$.
Thus, the 3-2 model predicts one massless active neutrino at low energy.
In the case of the 3-3 model, with three right-handed neutrinos, the Yukawa matrix is assumed to be rank two above an energy scale corresponding to the mass of the heaviest right-handed neutrino $N_3$.  
After we solve the renormalization group equations to energies below the lightest right-handed neutrino $N_1$, the rank of the effective mass matrix depends on the kernel solutions of the renormalization group equations.
For the simplest kernel solutions, the rank of the effective mass matrix does not change.
Meaning the lightest neutrino would remain massless across all renormalization energies, in spite of the different kernels.

We examine the reason for this result analytically. 
In general, at the energy scale higher than that of the heaviest right-handed neutrino mass, renormalization group equations do not change the rank of the effective Majorana mass matrix.
Similarly, in the energy region below the mass for the lightest right-handed neutrino, the renormalization group equations for the Wilson coefficient of the Weinberg operator do not the change its rank.
Therefore, the possible rank change may occur between the two energy regions where some right-handed neutrinos are still active and part of the effective Majorana mass matrix is described by the Wilson coefficient of the Weinberg operator.
For the 3-2 model, such energy region is unique ($M_1 <E< M_2$).
At high energy, the effective Majorana mass matrix is written with two linearly independent Yukawa vectors, which leads to the rank two effective mass matrix.
Effectively, the kernel solutions of the renormalization group equations act as transformations on the two linearly independent Yukawa vectors, and they change into another set of two linearly independent vectors.
Therefore, the effective Majorana mass matrix at low energy is also written with the two linearly independent Yukawa vectors leading to the rank two effective mass matrix.

In the framework of the 3-3 model with the rank degenerate Yukawa matrix, the effective mass matrix is the rank two at high energy scale, similar to the 3-2 model.
However, the form of the effective mass matrix is different since there are contributions from three Yukawa vectors.
Each of the three contributions is initially written in terms of three Yukawa vectors, and they are two linearly independent vectors and their superpositions.
These three contributions come from each right-handed neutrinos $N_1$-$N_3$.
The effect of renormalization group equations on the three vectors are different since there are two distinct energy regions $M_1 <E<M_2 $ and $M_2 <E<M_3$.
This leads to the rank of the effective mass matrix being dependent on the details of the renormalization group kernel solutions.
In the simplest case, the kernel solutions only depend on the neutrino Yukawa vectors.
Thus, the transformations of the kernel solutions can not increase the linearly vectors from two to three.
We note this may not be true for kernel solutions involving the charged lepton Yukawa vectors or for higher order renormalization group equations.
Both of those situations could be considered in a future analysis.

\section*{Acknowledgment}
The work of T.M. is supported by Japan Society for the Promotion of Science (JSPS) KAKENHI Grant Number JP17K05418.
N.J.B, would like to express thanks to the Japanese government Ministry of Education, Culture, Sports, Science, and Technology (MEXT) for the financial support during the writing of this work.

\appendix
\section{Renormalization Group Equations}\label{sec:RGE}
Here we list the one-loop renormalization group (RG) equations for the type-I seesaw models we are considering.
We simplified the RG equations by ignoring the Yukawa couplings for the quarks and charged the leptons.
Inclusion of the couplings does not change our results.
We start with the equations for the mass and coupling constant matrices unique to the type-I seesaw, $\overset{(n)}{M_R}(\mu)$, $\overset{(n)}{\kappa}(\mu)$, and $\overset{(n)}{y_{\nu}}(\mu)$;
\begin{gather}
   -16 \pi^2 \frac{d \overset{(n)}{\kappa}(t)}{dt} =
    \bigl(\overset{(n)}{y_{\nu}}(t)\overset{(n)}{y_{\nu}}(t)^{\dagger}\bigr) \kappa(t) +
    \overset{(n)}{\kappa}(t) \bigl(\overset{(n)}{y_{\nu}}(t)\overset{(n)}{y_{\nu}}(t)^{\dagger}\bigr)^T + \overset{(n)}{\alpha_{\kappa}}(t) \overset{(n)}{\kappa}(t),
    \label{eq:KappaRGE}  \\
  -16 \pi^2 \frac{d \overset{(n)}{y_{\nu}}(t)}{dt} =
    \left(\frac{3}{2}(\overset{(n)}{y_{\nu}}(t)\overset{(n)}{y_{\nu}}(t)^{\dagger}) +
    \overset{(n)}{\alpha_{y_{\nu}}}(t)\right)\overset{(n)}{y_{\nu}}(t),
    \label{eq:YukawaRGE} \\
  -16 \pi^2 \frac{d \overset{(n)}{M_R}(t)}{dt} =
    \bigl(\overset{(n)}{y_{\nu}}(t)^{\dagger}\overset{(n)}{y_{\nu}}(t)\bigr)^T
    \overset{(n)}{M_R}(t) + \overset{(n)}{M_R}(t)
    \bigl(\overset{(n)}{y_{\nu}}(t)^{\dagger}\overset{(n)}{y_{\nu}}(t)\bigr),
    \label{eq:RightMassRGE}
\end{gather}
where the scalar functions $\overset{(n)}{\alpha_{y_{\nu}}}(t)$ and $\overset{(n)}{\alpha_{\kappa}}(t)$ are given as;
\begin{gather}
  \overset{(n)}{\alpha_{y_{\nu}}}(t) \equiv 
    \Tr\bigl[\overset{(n)}{y_{\nu}}(t)\overset{(n)}{y_{\nu}}(t)^{\dagger}\bigr]-\frac{9}{20}g_1^2-\frac{9}{4}g_2^2, \\
  \overset{(n)}{\alpha_{\kappa}}(t) \equiv
    2\Tr\bigl[\overset{(n)}{y_{\nu}}(t)\overset{(n)}{y_{\nu}}(t)^{\dagger}\bigr]-3g_2^2-\lambda.
\end{gather}
In a manner similar to sections \ref{sec:32Seesaw} and \ref{sec:33Seesaw}, we use the superscript $n$ to notate the different energy regions.
The solutions to Eqs.(\ref{eq:KappaRGE}, \ref{eq:YukawaRGE}, \ref{eq:RightMassRGE}) are written in terms of a kernel function,
\begin{gather}
  \overset{(n)}{\kappa}(t_f) = e^{-\frac{1}{16\pi^2}
    \int_{t_i}^{t_f} \overset{(n)}{\alpha_\kappa}(s) ds}
  \overset{(n)}{W}(t_f-t_i) \overset{(n)}{\kappa}(t_i) \overset{(n)}{W}(t_f-t_i)^T,
    \label{eq:KappaRGESolution} \\
  \overset{(n)}{y_\nu}(t_f) = e^{-\frac{1}{16\pi^2} 
    \int_{t_i}^{t_f} \overset{(n)}{\alpha_{y_\nu}}(s) ds}
  \overset{(n)}{U}(t_f-t_i) \overset{(n)}{y_\nu}(t_i),
    \label{eq:YukawaRGESolution} \\
  \overset{(n)}{M_R}(t_f) =
\overset{(n)}{K}(t_f-t_i)  
  \overset{(n)}{M_R}(t_i) 
\overset{(n)}{K}(t_f-t_i)^T.
    \label{eq:RightMassRGESolution}
\end{gather}
The kernel functions are defined as,
\begin{align}
  \overset{(n)}{W}(t_f-t_i) & \equiv
  T \exp\left[-\frac{1}{16\pi^2}\int_{t_i}^{t_f}
  (\overset{(n)}{y_{\nu}}(s)\overset{(n)}{y_{\nu}}(s)^{\dagger}) ds \right],
      \label{eq:Kernelkappa} \\
  \overset{(n)}{U}(t_f-t_i) & \equiv
    T \exp\left[-\frac{1}{16\pi^2}\int_{t_i}^{t_f}\frac{3}{2}
    (\overset{(n)}{y_{\nu}}(s)\overset{(n)}{y_{\nu}}(s)^{\dagger}) ds \right],
      \label{eq:Kernelyukawa} \\
\overset{(n)}{K}(t_f-t_i) & \equiv  T \exp\left[-\frac{1}{16\pi^2}\int_{t_i}^{t_f}
    (\overset{(n)}{y_{\nu}}(s)^{\dagger}\overset{(n)}{y_{\nu}}(s))^T ds \right],
      \label{eq:KernelRightMass}
\end{align}
where, $T$ denotes the exponential ordered by decreasing energy scale.

\section{The proof of Eq.(\ref{eq:Eq.b1}) and Eq.(\ref{eq:Eq.bt1t2})} \label{sec:reverseKappaproof}
We prove the property of the kernel function for $\kappa$ on Yukawa matrix in Eq.(\ref{eq:Eq.b1}).
\begin{equation}
  \begin{split}
    \overset{(3)}{W}(t_2)\overset{(3)}{y_\nu}(0)&= T \exp\left[-\frac{1}{16\pi^2}\int_{0}^{t_2}(\overset{(3)}{y_{\nu}}(s)\overset{(3)}{y_{\nu}}(s)^{\dagger}) ds \right]\overset{(3)}{y_\nu}(0)
 \\
    &=\overset{(3)}{{y}_\nu}(t_2) T \exp\left[\frac{1}{32\pi^2}\int_{0}^{t_2}(\overset{(3)}{y_{\nu}}(s)^{\dagger} \overset{(3)}{y_{\nu}}(s)) ds \right]
e^{\frac{1}{16\pi^2}\int^{t_2}_0 \overset{(3)}{\alpha_{y_\nu}}(s) ds}. 
  \end{split}
\end{equation}
To prove the relation, we discretize the evolution kernel for $\kappa$ from Eq.(\ref{eq:Kernelkappa}) between the values $t_i=0$ and $t_f=N\epsilon$.
For which, $N$ is some arbitrary value larger than the discretization step $\epsilon$ and greater than zero,
\begin{multline}
  T\exp\left[-\frac{1}{16\pi^2}\int_{0}^{N \epsilon }(\overset{(3)}{y_{\nu}}(s)\overset{(3)}{y_{\nu}}(s)^{\dagger}) ds \right]
\overset{(3)}{y_{\nu}}(0)
\\
  =(1-\frac{1}{16 \pi^2}y[N-1]y[N-1]^\dagger \epsilon)\dotsm(1-\frac{1}{16 \pi^2}y[1]y[1]^\dagger \epsilon)(1-\frac{1}{16 \pi^2}y[0]y[0]^\dagger \epsilon ) y[0],
  \label{eq:discretizedKappa}
\end{multline}
where we denote $y[n] \equiv \overset{(3)}{y_{\nu}}(n\epsilon)$.
The discretization step $\epsilon$ is some small value, which approaches zero in the continuum limit.
Let us investigate the n th  step of the Yukawa coupling in Eq.(\ref{eq:discretizedKappa}) due to the kernel of kappa,
\begin{align}
  (1-\frac{1}{16 \pi^2}y[n]y[n]^\dagger \epsilon ) y[n]&=(1-\frac{3}{32 \pi^2}y[n]y[n]^\dagger \epsilon +\frac{1}{32 \pi^2}y[n]y[n]^\dagger \epsilon) y[n]
\\
  &= y[n+1](1+\frac{1}{32 \pi^2}y[n]^\dagger y[n] \epsilon) (1+\frac{\alpha_{y_\nu} [n]}{16 \pi^2} \epsilon),
  \label{eq:discretizedStep}
\end{align}
where $\alpha_{y_\nu} [n]\equiv \overset{(3)}{\alpha_{y_\nu}}(n \epsilon)$. Eq.(\ref{eq:discretizedStep}) is valid up to O($\epsilon$). To derive Eq.(\ref{eq:discretizedStep}), we use the descritized version
of the running of the Yukawa coupling of Eq.(\ref{eq:YukawaRGESolution} )
\begin{equation}
  (1+\frac{\alpha_{y_\nu} [n]}{16 \pi^2} \epsilon)y[n+1]=(1-\frac{3}{32 \pi^2}y[n]y[n]^\dagger \epsilon)y[n].
\end{equation}
This allows us to move the Yukawa coupling $y[n]$ from the right side to the left side. 
Applying Eq.(\ref{eq:discretizedStep}) on Eq.(\ref{eq:discretizedKappa}) multiple times leads to,
\begin{multline}
  (1-\frac{1}{16 \pi^2}y[N-1]y[N-1]^\dagger \epsilon ) \dotsm (1-\frac{1}{16 \pi^2}y[1]y[1]^\dagger \epsilon ) (1-\frac{1}{16 \pi^2}y[0]y[0]^\dagger \epsilon ) y[0]
\\
  = y[N](1+\frac{1}{32 \pi^2} y[N-1]^\dagger y[N-1] \epsilon ) \dotsm (1+\frac{1}{32 \pi^2} y[0]^\dagger y[0]\epsilon ) \prod_{n=0}^{N-1}  (1+\frac{\alpha_{y_\nu} [n]}{16 \pi^2} \epsilon).
\end{multline}
Lastly, we take the continuum limit to arrive at,
\begin{multline}
  T \exp\left[-\frac{1}{16\pi^2}\int_{0}^{t_2}(\overset{(3)}{y_{\nu}}(s)\overset{(3)}{y_{\nu}}(s)^{\dagger}) ds \right]\overset{(3)}{y_{\nu}}(0)
\\
  =\overset{(3)}{y_{\nu}}(t_2) T \exp\left[\frac{1}{32\pi^2}\int_{0}^{t_2}(\overset{(3)}{y_{\nu}}(s)^\dagger \overset{(3)}{y_{\nu}}(s)) ds \right] e^{\frac{1}{16\pi^2}\int^{t_2}_0 \overset{(3)}{\alpha_{y_\nu}}(s) ds}. 
  \label{eq:Eq.b2}
\end{multline}
One can also prove Eq.(\ref{eq:Eq.bt1t2}) similar to Eq.(\ref{eq:Eq.b1}), 
\begin{multline}
  T \exp\left[-\frac{1}{16\pi^2}\int_{t_2}^{t_1}(\overset{(2)}{y_{\nu}}(s)\overset{(2)}{y_{\nu}}(s)^{\dagger}) ds \right]\overset{(2)}{y_{\nu}}(t_2)
\\
  =\overset{(2)}{y_{\nu}}(t_1) T \exp\left[\frac{1}{32\pi^2}\int_{t_2}^{t_1}(\overset{(2)}{y_{\nu}}(s)^\dagger \overset{(2)}{y_{\nu}}(s)) ds \right] e^{\frac{1}{16\pi^2}\int^{t_1}_{t_2} \overset{(2)}{\alpha_{y_\nu}}(s) ds}.
 \label{eq:Eq.b2p}
\end{multline}
%
\vspace{0.2cm}
\noindent

\let\doi\relax

\end{document}